\documentstyle[12pt]{article}    
\begin{document}
\begin{center}
  
{\bf Exact Wavefunction in a Noncommutative Dipole Field Theory}

\vspace{1cm}

                      Wung-Hong Huang\\
                       Department of Physics\\
                       National Cheng Kung University\\
                       Tainan,70101,Taiwan\\

\end{center}
\vspace{3cm}

The nonrelativistic case of noncommutative scalar dipole field theory with quartic interaction on a two-dimensional spacetime is analyzed.    As there are two parameters in the general quartic interaction we try a way to find their relation.   To do this we first investigate the formulation of the quantum mechanics for a particle carrying the noncommutative dipole.   We point out a problem therein and propose a possible method to solve it.   We use this prescription to determined the quartic interaction term in the field theory once the noncommutative dipole is turned on.   The two particle Schr\"odinger equation is then solved and the exact wavefunction of the bound state and associated spectrum are found.  It is seen that the wavefunction has three center positions in the relative coordinates and the separation of the center is equal to the dipole length $L$ of the dipole field, exhibiting the nature of the noncommutative dipole field.   

\vspace{4cm}

\begin{flushleft}
   
E-mail:  whhwung@mail.ncku.edu.tw\\

\end{flushleft}

\newpage

   In recent a new type of noncommutative product is introduced by considering a system of $Dp$-branes in the presence of a constant background $B_{\mu\nu}$ field with one index along the branes' world volume and the other index transverse to it [1,2]. The field theories defined on these $Dp$-branes are also noncommutative, but, in contrast to the Moyal case [3], the space-time remains commutative. The noncommutativity appears only in the product of functions and has its origin in the finite dipole length $\vec{L}$ associated to each field. A supergravity description of these so called noncommutative dipole theories is
presented in [4].

    In this paper we will consider the nonrelativistic case of noncommutative scalar dipole field theory with the quartic interaction on two-dimensional spacetime.   As there are two parameters in the general quartic interaction we try to search a principle to find their relation.   To do this we first investigate the formulation of the quantum mechanics for a particle carrying the noncommutative dipole.   It is found that there is a problem in formulizing the quantum mechanics when the wavefunction carrying the internal noncommutative dipole.  We propose a possible method to solve this problem.   Using the prescription we see that, the ground state wavefunction for a particle possessing a noncommutative dipole in the simple harmonic potential will  have two centers, which correctly reveals the nature of the noncommutative dipole in the particle.

 Using this prescription we then determine the interaction term in the field theory once the noncommutative dipole is turned on and solve the two particle Schr\"odinger equation.   We find the exact wavefunction of the bound state.  The wavefunction shows three center positions in the relative coordinates and the separation of the center is equal to the dipole length $\vec{L_i}$ of the dipole field.   This clearly exhibits the dipole nature of the noncommutative dipole field theory.

    Note that the three-dimensional spacetime noncommutative field theory (in the Moyal case) has been investigated in [5].   The simplest space in the Moyal case is two  dimensions while that in the noncommutative scalar dipole system could be only one dimension and does not need  renormalization. 

  Let us begin with a brief summary about the basic property of the algebra of noncommutative dipole field.  The noncommutative dipole field theory was first studied in [6] and then in [7,8].

  First, as in the Moyal case, the noncommutative dipole field theory can also be defined by replacing the ordinary product of function by a noncommutative dipole $\star$-product.   To any dipole field, $\phi_i$, we assign a {\it constant} space-like dipole length $\vec{L_i}$ and define the ``dipole star product'' as

$$ (\phi_i * \phi_j)(x) \equiv \phi_i(x - {1\over 2} L_j)~ \phi_j (x +
{1\over 2}L_i). \eqno{(1)}$$

   Next, the natural choice for the $Tr$, similar to the Moyal case, is the integral over the space-time. Demanding the integral over star products of arbitrary dipole field to enjoy the cyclicity condition we shall restrict the integrand to have a total zero dipole length. This means that the following integral 

$$\int \phi_1 *\phi_2 *\cdots * \phi_n = \int \phi_n* \phi_1 *\cdots
\phi_{n-1} , \eqno{(2)} $$
serves as the proper $Tr$, only if the condition 
$$\sum_{i=1}^n \vec{L_i}=0, \eqno{(3)}$$
is satisfied, where $\vec L_i$ are the dipole lengths for $\phi_i$.  In the field theory this condition is translated into the fact that any term in the proposed action should have a total vanishing dipole length. Therefore in each vertex, both the sum of the external momenta and the total dipole length should vanish [6].

  Finally, we also need to define complex conjugate of a field and the behavior of  the star product under complex conjugation.  Demanding $(\phi^{\dagger}*\phi)$ to be real valued, i.e. $\phi^\dagger *\phi=(\phi^\dagger\ *\phi)^{\dagger}$, we can fixes the dipole length of $\phi^\dagger$ to be the same as that of $\phi$ though with a minus sign. Therefore the dipole length of any real (hermitian) field, and in particular the gauge fields, is zero [6].  Not that the noncommutativity in the dipole case is not a property of space-time, but originating from the dipole length associated to each field.   Thus, the derivative operator on the dipole field will behave like the usual derivative [6].

   Along the above property, the action for the quartic interaction scalar
noncommutative dipole theory is

$$S=\int \left[\partial_\mu\phi^\dagger * \partial_\mu\phi - V_*(\phi^\dagger*\phi) \right], \eqno{(4)}$$
where $V_*(\phi^\dagger*\phi)$ is the potential with the products replaced by star products in (1).   Using the cyclicity condition, we can drop the star product in the quadratic part of the action (much like the Moyal-star product case) and the effects of dipole moments appear only through the interaction terms.   The most general case of quartic interaction  $V_*(\phi^\dagger*\phi)$ is [6]

$$ V_*(\phi^\dagger*\phi) = \lambda_0(\phi^\dagger *\phi) *(\phi^\dagger *\phi) +2\lambda_1(\phi^\dagger *\phi) *(\phi *\phi^\dagger), \eqno{(5a)}$$
and therefore
$$\int V_*(\phi^\dagger*\phi)=\int \left[\lambda_0 (\phi^\dagger \phi)(x)(\phi^\dagger \phi)(x) + 2\lambda_1 (\phi^\dagger \phi)(x - {L\over 2})(\phi^\dagger \phi)(x + {L\over 2})\right] \hspace{3cm}$$
$$=\int \left[\lambda_0 (\phi^\dagger \phi)(x)(\phi^\dagger \phi)(x) + \lambda_1 (\phi^\dagger \phi)(x)(\phi^\dagger \phi)(x +{L}) +\lambda_1 (\phi^\dagger \phi)(x - L)(\phi^\dagger \phi)(x)\right]. \eqno{(5b)}$$
The renormalization of the above noncommutative dipole theory has been discussed in [6].  It was shown that the theory, though non-local, would
be renormalizable in the usual sense. 

   Note that the most general quartic interaction  $V_*(\phi^\dagger*\phi)$ described in (5a) has two parameter $\lambda_0$ and $\lambda_1$.   These two parameters seem to be irrelevant to each other and shall be defined by the model system.   However, the following investigation could help us to determinate the relation between these two parameters.

   To do this let us consider a quantum field without a noncommutative dipole.   In this case the interaction may be written as   $V =  \lambda (\phi^\dagger \phi) (\phi^\dagger \phi) = {1\over 2}\lambda (\phi^\dagger \phi) (\phi^\dagger \phi) + {1\over 2} \lambda (\phi^\dagger \phi) (\phi \phi^\dagger)$.   Therefore, after turning on the noncommutative dipole we can expressed the interaction as that in (5a).   Thus, we see that there are the following relations: $\lambda_0 ={1\over 2} \lambda $ and $\lambda_1 ={1\over 4} \lambda $.    In this way, the interaction term in a noncommutative dipole field theory can be determined from the corresponding commutative theory. 

  To investigate this problem furthermore let us investigate the quantum mechanics system.   As in the Moyal case [9,10],  the general action of  the quantum mechanics for a particle carrying the noncommutative dipole can be expressed as

$$ S=\int dt dx {\bar\Psi(x,t)}*\left[ i {\partial \over \partial t}- {p^2 \over 2m}-V(x) \right]*\Psi(x,t)$$
$$ = \int dt dx {\bar\Psi(x,t)}\left[ i {\partial \over \partial t}- {p^2 \over 2m}-V(x-{L\over 2}) \right] \Psi (x,t). \eqno{(6.a)}   $$
The above relation tell us that when a conventional particle moving in a potential $V(x)$, then after turning on its noncommutative dipole the new particle will  behave as a conventional one while under the potential $V(x-{L\over 2})$.   Therefore, the particle spectrum will not be changed even if its noncommutative dipole was turned on and the associated wavefunction is just a conventional particle while shifting $x\rightarrow x+{L\over 2}$.    However, this simple property suffers a serious problem.   For example,  consider a system which has a potential $V(x)=0, $ as $|x| \le \ell$ and  $V(x)= \infty, $ as $|x| \ge \ell$.   (This potential can be regarded as a limit form of some continuous functions.)  In this case, according to the above discussion, a particle with noncommutative dipole $|L|$ could appear at the region $\ell \le x \le \ell +{|L|\over 2}$.  But, as an intuition, because the potential in this region is infinity and particle shall not be found in this region.  Therefore, there are some problems in the action of (6.a).   

   To solved this problem we propose that the action in (6.a) shall be replaced by

$$ S=\int dt dx {\bar\Psi(x,t)}* \left[i {\partial \over \partial t}- {p^2 \over 2m}\right]*\Psi(x,t) \hspace{6cm}$$
$$- \int dt dx {1\over 2}\left[{\bar\Psi(x,t)}*V(x) *\Psi(x,t) + \Psi(x,t)* V(x)*{\bar\Psi(x,t)} \right]$$
$$ = \int dt dx {\bar\Psi(x,t)}\left[ i {\partial \over \partial t}- {p^2 \over 2m}-{1\over 2}[~V(x-{L\over 2}) + V(x+{L\over 2})~]\right] \Psi (x,t). \eqno{(6.b)}   $$
From the above relation we see that a conventional  particle confined in the region $|x| \le \ell$ will be confined in the new region $|x| \le \ell-|{L\over2}|$ once the noncommutative dipole is turned on.   Especially, it is interesting to consider a conventional  particle moving in the potential $V={1\over 2} k x^2$.    The ground state of this system has one center at $x=0$.   However, when noncommutative dipole of the particle is turned on, the new particle will behave as a conventional  particle moving in the potential $V={1\over 4} k (x-{L\over 2})^2 + {1\over 4} k (x+{L\over 2})^2$.    The ground state of this new particle, which possesses a noncommutative dipole $L$, will  have two centers at $x={L\over 2}$ and $x=-{L\over 2}$.  Therefore, the wavefunction now correctly reveal the nature of the noncommutative dipole in the particle.

   We thus conjecture that the interaction term in the noncommutative dipole field theory shall be the summations of all possible permutations of the field in the interaction of commutative field theory, and replace the ordinary product by the dipole star product defined in (1).

   We now consider the nonrelativistic limit of the noncommutative dipole scalar field theory (4) on one space described by the Lagrangian,

$$ L=\int dt dx \left[ i\,\phi^\dagger \partial_t \phi + {1\over 2} \phi^\dagger  \partial_x ^2 \phi - V_*(\phi^\dagger*\phi) \right], \eqno{(7)}$$
in which $V_*(\phi^\dagger*\phi) $ is defined in (5b) while using the relations $\lambda_0 ={1\over 2} \lambda $ and $\lambda_1 ={1\over 4} \lambda $.  Following the method in [5] we quantize the system by the canonical quantization method imposing the canonical commutation 
relation  

$$ [\phi(x),\phi^\dagger(y)]=\delta(x-y), \eqno{(8)}$$
The Hamiltonian now is given by 

$$H=\int dt dx\left[ -{1\over 2} \phi^\dagger \partial_x ^2
 \phi  + V_*(\phi^\dagger*\phi) \right]. \eqno{(9)}$$
Note that the interaction $V_*$ in here represents a contact interaction term in the usual sense.   

   We now construct two particle Schr\"odinger equation by the following manner [5]. The operator Schr\"odinger equation is described by

$$i {\partial\phi(x)\over \partial t} = [\phi(x), H] = - {1\over 2}\partial_x^2 \phi(x)+ \lambda \phi^\dagger(x)\phi(x)\phi(x) +\hspace{3cm}$$
$$ {\lambda\over 2} \phi(x)\phi^\dagger(x+L)\phi(x+L) + {\lambda\over 2}\phi(x)\phi^\dagger(x-L)\phi(x-L) ,  \eqno{(10)}$$
where the time argument of the Schr\"odinger field operator is suppressed for simplicity. The two particle wavefunction may be constructed by projecting a generic state $|\Phi\rangle$ to two particle sector, i.e. $\Phi(x,y) =\langle 0|\phi(x)*\phi(y)|\Phi\rangle$. Then, using the operator Schr\"odinger equation we can find the two particle Schr\"odinger equation,

$$i \partial_ t\Phi(x,y) = - {1\over 2}(\partial_x^2 + \partial_y^2)\Phi(x,y)+ 2\lambda \delta (x-y-L)\Phi(x,x-L)$$
$$ \hspace{5cm}+ \lambda \delta(x-y)\Phi(x,x) +\lambda \delta(x-y-2L)\Phi(x,x-2L) \eqno{(11)}$$
Defining the momentum-space wavefunction

$$\Phi(Q, q)=\int dx dy~ e^{-{i\over 2}Q(x+y)-i q(x-y)} \Phi(x,y),\eqno{(12)}$$
equation (11) in the momentum space becomes

$$i\dot{\Phi}(Q,q) = \left({1\over 4}{Q^2}+q^2\right) \Phi(Q,q) + \lambda \int d\tilde q ~\Phi (Q, \tilde q)+\hspace{4cm}$$
$$+ 2 \lambda \int d\tilde q ~e^{-i (\tilde q - q)2L}~\Phi (Q, \tilde q)+ \lambda \int d\tilde q ~~e^{-i (\tilde q - q)L}~\Phi (Q, \tilde q).\eqno{(13)}$$
This equation is already diagonal with respect to $Q$ and after setting
 
$$\Phi(Q,q)= \delta(Q-P)~\Psi(P,q)~e^{-i({1\over 4 }{P^2}+E_r)t},\eqno{(14)} $$
the Schr\"odinger equation is reduced to 

$$ (E_r-q^2)\Psi(P,q)=  \lambda \int d\tilde q ~\Psi (Q, \tilde q)+ 2 \lambda \int d\tilde q ~e^{-i (\tilde q - q)2L}~\Psi (Q, \tilde q) $$
$$ +  \lambda \int d\tilde q ~~e^{-i (\tilde q - q)L}~\Psi (Q, \tilde q).  \eqno{(15)}$$
Now, consider the bound state with $E_r= -\epsilon_B$, equation (15) becomes

$$\Psi(P,q)= {-1\over q^2+ \epsilon_B}[~  \lambda \int d\tilde q ~\Psi (Q, \tilde q) + 2 \lambda \int d\tilde q ~e^{-i (\tilde q - q)2L}~\Psi (Q, \tilde q)$$
$$~~+ \lambda \int d\tilde q ~~e^{-i (\tilde q - q)L}~\Psi (Q, \tilde q)~].\eqno{(16)}$$
The above equation can be solved exactly.     We first define   

   $$\Psi(P,q)= \int dR~dr~ e^{iRP} e^{irq}~\Psi(R,r).\eqno{(17)}$$
Then, after integrating the both sides for $q$ with the weight factor  $e^{irq}$, equation (16) becomes 

  $$\Psi(R,r)= e^{iRP}\psi(r),\hspace{10cm}\eqno{(18)}$$
with
$$\psi(r)= - \lambda {\pi\over {\sqrt \epsilon_B}} e^{-\mid r\mid{\sqrt \epsilon_B}}\psi(0) - 2 \lambda  {\pi\over {\sqrt \epsilon_B}} e^{-\mid r-L\mid{\sqrt \epsilon_B}}\psi(L) -  \lambda {\pi\over {\sqrt \epsilon_B}} e^{-\mid r-2L\mid{\sqrt \epsilon_B}}\psi(0),\eqno{(19)}$$
where $ R= {x + y\over 2}$ and the relative position $r$ denotes $x-y$.  We thus get the explicit form of the position space wavefunction.     It is interesting to see that the wavefunction in (19) is simply an addition of three function which coming from different interaction in (5b).    To find the spectrum of the two particle bound state, we can set $r=0$ and $r=L$ respectively in (19) and then solve the equations.   The eigenvalue equation for the energy  $\epsilon_B$ is 

$${{\sqrt \epsilon_B}+ 2\lambda \pi \over  2\lambda \pi~e^{-L{\sqrt \epsilon_B}}} = {2\lambda \pi \over  {\sqrt \epsilon_B}+ \lambda \pi (1+ e^{-2L{\sqrt \epsilon_B}})}. \eqno{(20)} $$
From the above results we see that in the commutative case, $L=0$, there is a possible bounded state with the spectrum ${\sqrt \epsilon_B}= - 4\lambda \pi$ and wavefunction has only single center.   This corresponds to the two particle system interacting through an attractive delta force (if $\lambda<0$).  In general there are three distinguished centers at $r=0, L$ and $2L $ in the relative coordinates and we could have a bound state if $\lambda<0$.  This behavior signaling the nature of the field theories with noncommutative dipole length.   

   In conclusion, we have proposed a method to formulate the action of a quantum mechanics when the internal noncommutative dipole was turned on.  In this way we can avoid some problems and wavefunction could then correctly reveal the nature of the noncommutative dipole in the particle.  Note that the  action we propose in  (6.b) is different from that in the Moyal case (6.a).   We use our prescription to fix the quartic interaction in a noncommutative scalar dipole field theory.  The two particle Schr\"odinger equation on a two-dimensional spacetime is then solved and the exact wavefunction of the bound state and associated spectrum are found.  We see that the wavefunction has three center positions in the relative coordinate which exhibiting the nature of the noncommutative dipole field.

\newpage

\begin{enumerate}

\item   A. Bergman, and O. J. Ganor, {Dipoles, Twists and Noncommutative
   Gauge Theory}, JHEP {\bf 0010} (2000) 018, hep-th/0008030.
\item  A. Bergman, K. Dasgupta, O. J. Ganor, J. L. Karczmarek, and G.
   Rajesh, {Nonlocal Field Theories and their Gravity Duals}, Phys. Rev. {\bf D65} (2002) 066005, hep-th/0103090.
\item    M. R. Douglas, and N. A. Nekrasov, {Noncommutative Field
   Theory}, Rev. Mod. Phys. {\bf 73} (2002) 977, hep-th/0106048, and references therein.
\item    M. Alishahiha, and H. Yavartanoo, {Supergravity Description of
  the Large \emph{N} Noncommutative Dipole Field Theories}, JHEP {\bf 0204} (2002) 031, hep-th/0202131.
\item  D. Bak, S. K. Kim, K. S. Soh and J. H. Yee,  {Exact Wavefunctions in a Noncommutative Field Theory}, Phys. Rev. Lett. {\bf 85} (2000) 3087, hep-th/0005253; {Noncommutative Field Theories and Smooth Commutative Limits}, Phys.Rev. {\bf D63} (2001) 047701, hep-th/0006087.
\item    K. Dasgupta, and M. M. Sheikh-Jabbari, {Noncommutative Dipole Field Theories}, JHEP {\bf 0202} (2002) 002, hep-th/0112064.
\item   D. H. Correa, G. S. Lozano, E. F. Moreno, and F. A.
   Schaposnik, {Anomalies in Noncommutative Dipole Field
   Theories}, JHEP {\bf 0202} (2002) 031, hep-th/0202040.
\item   N. Sadooghi, and M. Soroush, {\it Noncommutative Dipole QED}, hep-th/0206009.

\item  M.~Chaichian, M.~M.~Sheikh-Jabbari and A.~Tureanu, ``Hydrogen atom spectrum and the Lamb shift in noncommutative QED,'' Phys.\ Rev.\ Lett.\  {\bf 86}, 2716 (2001), hep-th/0010175; D.~Kochan and M.~Demetrian, ``QM on non-commutative plane,'' Acta Physica Slovaca 52, No. 1, (2002), 1, hep-th/0102050; 
Pei-Ming Ho and Hsien-Chung Kao, ``Noncommutative Quantum Mechanics from Noncommutative Quantum Field Theory '', Phys.\ Rev.\ Lett.\  {\bf 88}, 151602 (2002), hep-th/0110191.

\item  J.~Gamboa, M.~Loewe and J.~C.~Rojas, ``Non-Commutative Quantum Mechanics,'' Phys.\ Rev.\ D {\bf 64}, 067901 (2001), hep-th/0010220; V.~P.~Nair and A.~P.~Polychronakos, ``Quantum mechanics on the noncommutative plane and sphere,'' Phys.\ Lett.\ B {\bf 505}, 267 (2001), hep-th/0011172; B.~Morariu and A.~P.~Polychronakos, ``Quantum mechanics on the noncommutative torus,''
Nucl.\ Phys.\ B {\bf 610}, 531 (2001), hep-th/0102157.

\end{enumerate}
\end{document}